\documentclass[sigconf,screen]{acmart}

\usepackage{booktabs}
\usepackage{multirow}
\usepackage{tabularx}
\usepackage{algorithm}
\usepackage{algpseudocode}

\usepackage{amsmath,amssymb}
\usepackage{xspace}
\usepackage{enumitem}
\usepackage{graphicx}
\usepackage{array}
\usepackage{float}
\usepackage{placeins}
\usepackage{microtype}

\newcommand{\singletablewidth}{0.96\linewidth}
\copyrightyear{2026}
\acmYear{2026}
\setcopyright{cc}
\setcctype{by-nc-nd}
\acmConference[ASE '26]{Proceedings of the 41st IEEE/ACM International Conference on Automated Software Engineering}{October 12--16, 2026}{Munich, Germany}
\acmBooktitle{Proceedings of the 41st IEEE/ACM International Conference on Automated Software Engineering (ASE '26), October 12--16, 2026, Munich, Germany}
\acmDOI{10.1145/3832783.3837440}
\acmISBN{979-8-4007-2882-2/2026/10}
\acmSubmissionID{ase26main-p423-p}
\received{2026-03-26}
\received[accepted]{2026-06-18}

\begin{document}

\newcommand{\sys}{\textsc{TELLER}\xspace}
\newcommand{\tpe}{\textsc{TPE}\xspace}
\newcommand{\mot}{\textsc{MoT}\xspace}

\title[TELLER: Non-intrusive Cross-Layer Root-Cause Analysis for LLM Inference]{TELLER:\texorpdfstring{\allowbreak\ }{ }\texorpdfstring{Non\mbox{-}intrusive}{Non-intrusive} Cross-Layer Root-Cause Analysis for \texorpdfstring{\mbox{LLM Inference}}{LLM Inference}}

\makeatletter
\patchcmd{\@mkbibcitation}
  {In \textit{\@acmBooktitle}}
  {In \textit{\@acmBooktitle.}}
  {}{}
\patchcmd{\@mkbibcitation}
  {\ifx\@acmEditors\@empty\textit{.}\else}
  {\ifx\@acmEditors\@empty\else}
  {}{}
\pretocmd{\@mkbibcitation}{%
  \def\@title{TELLER:\linebreak \texorpdfstring{Non\mbox{-}intrusive}{Non-intrusive} Cross-Layer Root-Cause Analysis for \texorpdfstring{\mbox{LLM Inference}}{LLM Inference}}%
}{}{}
\makeatother

\author{Ruilin Xu}
\authornote{These authors contributed equally to this work.}
\orcid{0009-0006-8235-3345}
\email{xurlin5@mail2.sysu.edu.cn}
\affiliation{%
  \institution{Sun Yat-sen University}
  \city{Guangzhou}
  \country{China}
}
\author{Junyi Li}
\authornotemark[1]
\orcid{0009-0004-7501-2142}
\email{lijy727@mail2.sysu.edu.cn}
\affiliation{%
  \institution{Sun Yat-sen University}
  \city{Guangzhou}
  \country{China}
}
\author{Pengfei Chen}
\authornote{Corresponding authors.}
\correspondingauthor
\orcid{0000-0003-0972-6900}
\email{chenpf7@mail.sysu.edu.cn}
\affiliation{%
  \institution{Sun Yat-sen University}
  \city{Guangzhou}
  \country{China}
}
\author{Zongxuan Xie}
\authornotemark[2]
\orcid{0009-0000-4067-5668}
\email{xiezx25@mail2.sysu.edu.cn}
\affiliation{%
  \institution{Sun Yat-sen University}
  \city{Guangzhou}
  \country{China}
}

\renewcommand{\shortauthors}{R. Xu, J. Li, P. Chen, and Z. Xie}

\begin{abstract}
Large language model (LLM) inference has evolved from an offline workload into a continuously operated software service, yet root-cause analysis remains difficult because a single request spans the inference engine, Python/C++ backend, host CUDA APIs, GPU kernels, and distributed communication. Existing profilers expose raw timelines, while log-based diagnosis often misses cross-layer execution semantics and request-level structure. We present \sys, a non-intrusive \textbf{T}race- and \textbf{L}og-aware \textbf{L}LM inf\textbf{E}rence \textbf{R}oot-cause analysis framework. \sys first collects NVTX/CUPTI traces and service logs without modifying model binaries, then reconstructs per-request call-chain trees and aligns log lines with the corresponding execution steps. We introduce a dependency-aware causal-context slice that preserves parent--child structure, temporal order, and communication relations, and a Trace Pair Encoding (\tpe) tokenizer that compresses such slices into compact structural token sequences with parent, depth, and duration attributes. On top of these representations, \sys combines numeric candidate localization with a multimodal root-cause model that jointly predicts abnormal steps, localizes suspicious operators, and generates natural-language explanations. Experiments on multi-node GPU inference workloads show a clear compression--accuracy trade-off: a moderate \tpe vocabulary reduces per-step trace length by more than 80\% while achieving the best overall performance on both horizontal (cross-node communication) and vertical (within-node execution stack) views, whereas more aggressive compression substantially degrades diagnosis quality. Further analyses under low-fault priors, strengthened baselines, modality ablations, explanation-quality checks, and tracing overhead show that \sys provides a practical triage and evidence-localization substrate for LLM inference RCA.
\end{abstract}




\maketitle

\section{Introduction}
Large language models (LLMs) are increasingly deployed as always-on software services.
Modern inference systems operate under sustained concurrency, heterogeneous accelerators, and distributed communication, and therefore exhibit a deep software--hardware stack that spans the inference engine, tensor runtime, CUDA host APIs, GPU kernels, and network collectives~\citep{brown2020gpt3,touvron2023llama,bai2023qwen,kwon2023pagedattention,paszke2019pytorch,dao2022flashattention}.
When a production incident occurs, the practical question is rarely whether a request is merely abnormal.
Instead, engineers need a root cause: \emph{what failed, where it failed, when the deviation first became visible, and which operators or communication steps were involved?}
This is precisely the kind of automated debugging and diagnosis problem that modern software engineering should address.

However, root-cause analysis (RCA) for LLM inference remains difficult.
First, a request is not executed as a simple sequential program.
Continuous batching, coroutine scheduling, asynchronous CUDA launches, and overlapping communication create many-to-one mappings between requests and low-level operators.
Second, the observability signals themselves are heterogeneous.
Traces provide precise timing and dependency information, but they are low-level and voluminous, while logs carry semantic hints and error descriptions, but they are sparse and often detached from the exact operator or kernel that triggered the failure.
Third, current practice remains tool-fragmented.
Timeline profilers help experts inspect kernels and runtime calls, whereas log-based tools reason over textual symptoms, yet neither alone offers a compact, request-level, cross-layer explanation.
Finally, production serving imposes strict operational constraints: instrumentation should be non-intrusive, lightweight, and independent of model binaries.
We therefore position \sys as an initial triage and evidence-localization filter for engineers: it narrows a large trace/log space to likely faulty steps, operators, and explanations, while later human or domain-specific validation remains necessary for rare, ambiguous, or high-impact incidents.

The broader systems community has long recognized the importance of low-overhead, request-centric tracing in large services, from Pinpoint to Dapper and later observability stacks~\citep{chen2002pinpoint,sigelman2010dapper,opentelemetry2024}.
Yet LLM inference introduces a richer vertical stack and much denser asynchronous execution than classic web-service request processing.
Similarly, existing log-based or LLM-based diagnosis methods demonstrate strong textual reasoning ability, but they often ignore GPU kernels, CUDA launches, and communication patterns that are central to inference failures~\citep{zhang2019robustlog,lee2021lanobert,li2023logs2graphs,li2019madgan,ahmed2023rootcause,chen2024auto,jin2023assess,roy2024agents,wang2024netassistant}.

These challenges motivate a new diagnosis substrate for LLM serving.
Rather than treating traces as visualization output or treating logs as the sole diagnosis input, we argue that RCA should be built on a unified trace--log representation that preserves execution structure, timing, and semantic context at the request level.
To address this need, we propose \sys, a non-intrusive \textbf{T}race- and \textbf{L}og-aware \textbf{L}LM inf\textbf{E}rence \textbf{R}oot-cause analysis framework.

The key idea behind \sys is to separate diagnosis into two stages.
A lightweight localization stage operates on merged cross-layer traces and narrows the search space to suspicious operator families and their neighborhoods.
A multimodal reasoning stage then explains the failure using a compact, dependency-aware representation of the suspicious region together with aligned logs.
To make this feasible for downstream models, \sys introduces two representation techniques.
First, it extracts a \emph{causal-context slice}: a compact subgraph around suspicious operator instances enriched with structural ancestry, temporal succession, and communication coupling.
Second, it introduces \emph{Trace Pair Encoding} (\tpe), a structure-preserving tokenizer that compresses trace slices into symbolic tokens while maintaining parent indices, durations, and depths.

In summary, this paper makes four main contributions.

\noindent\textbf{(1) Non-intrusive cross-layer trace--log diagnosis.}
We present an end-to-end pipeline that collects NVTX/CUPTI traces and service logs, reconstructs per-request call-chain trees, and aligns logs with the corresponding execution steps without modifying model binaries.

\noindent\textbf{(2) A dependency-aware causal-context representation.}
We introduce a causal-context slice that compactly summarizes suspicious regions by preserving parent--child structure, temporal order, and communication relations, thereby answering who, where, when, and with whom a failure is associated.

\noindent\textbf{(3) Structure-preserving trace compression and multimodal RCA.}
We propose \tpe, a trace-specific structural tokenizer that maintains parent, duration, and depth information under BPE-style merges, and combine it with a structure- and time-aware Trace Encoder and a joint log--trace root-cause model.

\noindent\textbf{(4) Extensive experimental evaluation.}
We conduct experiments covering step-level detection, operator-level localization, explanation quality, low-fault-prior behavior, modality contribution, tracing overhead, and strengthened structured-feature baselines. The results show that \sys provides strong diagnosis and localization quality as an RCA triage substrate, while highlighting the importance of moderate structure-preserving compression.

\section{Background and Problem Definition}\label{sec:background}

\subsection{LLM Inference as a Cross-Layer Software Stack}
A typical LLM inference deployment can be abstracted into four tightly coupled layers~\citep{kwon2023pagedattention,paszke2019pytorch,nvidia2024cuda}.
The \emph{inference engine layer} schedules requests, manages KV cache, and forms execution steps such as prefill and decoding.
The \emph{compute backend layer} maps high-level tensor operations to framework- and library-level implementations.
The \emph{host-operator layer} launches CUDA runtime and driver calls, manages streams and events, and orchestrates memory movement.
The \emph{device-operator layer} executes GPU kernels and memory operations.
In distributed settings, communication libraries such as NCCL introduce a horizontal dimension through collectives, synchronization, and transport behavior.

Two systems properties dominate the diagnosis problem.
\emph{Asynchrony} means that host calls and device kernels are not naturally aligned on a single timeline.
\emph{Concurrency} means that the engine merges multiple requests into shared operators or shared communication groups.
As a result, raw events cannot be interpreted locally, and they must be reconciled into request-level structures that preserve both temporal and semantic information.

\subsection{From Observability to Root-Cause Analysis}
Existing observability tools cover only part of the problem.
Profilers such as Nsight Systems and PyTorch Profiler expose operator timelines, kernels, and API activity, but they are fundamentally low-level analysis tools and do not directly return request-level root causes~\citep{nvidia2025nsight,pytorch2025profiler,tensorflow2020profiler}.
Classical anomaly detectors such as KMeans, DBSCAN, Isolation Forest, and GMM provide stable numeric signals but are weak at semantic interpretation~\citep{hartigan1979kmeans,ester1996dbscan,liu2008isolationforest,reynolds2009gmm,xu2025eacgm}.
Recent log-based or LLM-based diagnosis systems can produce human-readable explanations, yet they usually reason over textual symptoms rather than over GPU kernels, host launches, and communication structure~\citep{zhang2019robustlog,lee2021lanobert,li2023logs2graphs,li2019madgan,ahmed2023rootcause,roy2024agents}.
LLM inference RCA therefore demands a unified representation that connects
(i) low-level execution evidence,
(ii) structural execution context, and
(iii) semantic diagnosis cues.

\subsection{Problem Statement}
Let \(q\) denote a request and \(s\) one execution step of that request.
After cross-layer merging, \sys reconstructs a request-level call-chain tree
\[
\mathcal{C}_q(s)=\bigl(V_q(s),E_q(s)\bigr),
\]
whose nodes cover step ranges, front-end and back-end operators, host API calls, device kernels, and communication events.
From the same run, we also collect an aligned log snippet
\[
\mathcal{L}_q(s)=\{\ell_1,\ldots,\ell_m\},
\]
where each \(\ell\) contains at least a timestamp and message content, and may also contain request identifiers and log levels.

The RCA input is thus
\[
x_s = \bigl(\mathcal{C}_q(s), \mathcal{L}_q(s)\bigr).
\]
The output of interest is a triplet
\[
f_\theta(x_s)=\bigl(\hat{y}_s,\hat{r}_s,\hat{\mathcal{O}}_s\bigr),
\]
where \(\hat{y}_s\) is a step-level abnormality decision or fault label,
\(\hat{r}_s\in\Sigma^\ast\) is a natural-language root-cause explanation,
and \(\hat{\mathcal{O}}_s\subseteq V_q(s)\) is an evidence set of suspicious operators or communication events.

We target four requirements:
\emph{(R1)} request-level attribution despite batching and asynchrony;
\emph{(R2)} compact summaries rather than full raw traces;
\emph{(R3)} non-intrusive deployment without binary modification;
and \emph{(R4)} transferability across engines, workloads, and fault categories.

\begin{figure*}[h!]
\centering
\includegraphics[width=0.85\textwidth,alt={A cross-layer LLM inference trace that connects application operators, CUDA calls, GPU kernels, and communication events to TELLER's diagnosis output.}]{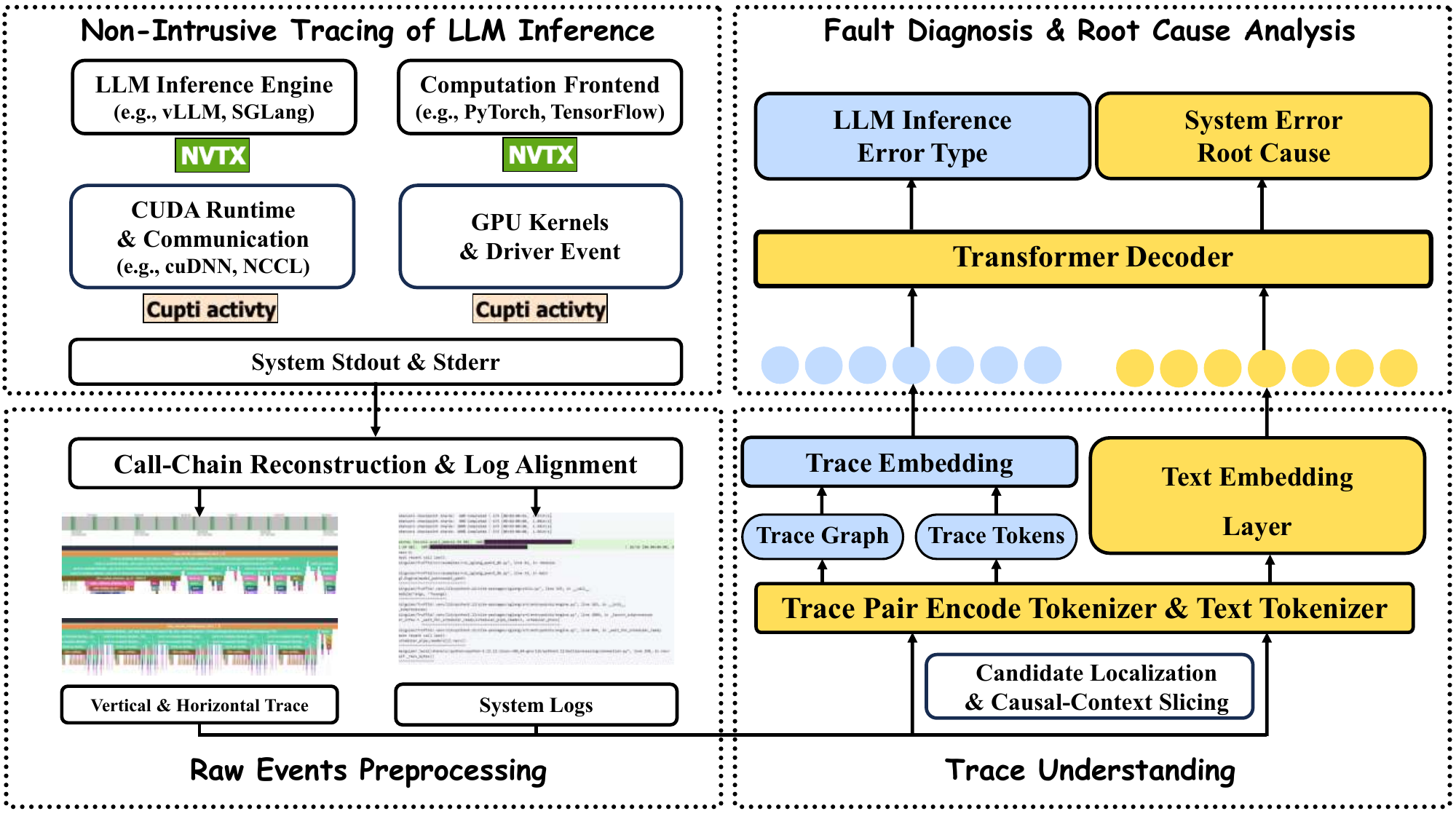}
\caption{Overview of \sys. \sys first performs non-intrusive cross-layer tracing of LLM inference, then reconstructs request-level call chains with aligned logs, extracts compact suspicious causal contexts, and finally conducts multimodal fault diagnosis and root-cause analysis to predict error types and generate root-cause explanations.}
\Description{An end-to-end TELLER workflow. Cross-layer trace and log collection feed request-level call-chain reconstruction and log alignment, followed by candidate localization, causal-context slicing, TPE encoding, and multimodal root-cause diagnosis.}
\label{fig:overview}
\end{figure*}

\section{\sys}\label{sec:method}

\subsection{Overview}
Figure~\ref{fig:overview} illustrates the end-to-end workflow of \sys.
At a high level, \sys consists of two stages.
The first stage transforms raw runtime signals into aligned diagnosis-ready evidence by performing non-intrusive tracing of LLM inference and raw-events preprocessing, including request-level call-chain reconstruction and log alignment.
The second stage transforms the reconstructed evidence into diagnosis outputs through trace understanding and fault diagnosis, including candidate localization, causal-context slicing, compact trace--log encoding, error-type prediction, and root-cause generation.

Concretely, the pipeline proceeds in four steps.
First, \sys performs non-intrusive cross-layer collection by injecting NVTX- and CUPTI-based tracing into the LLM inference engine, computation frontend, CUDA runtime, communication library, GPU driver, and kernel execution, while simultaneously capturing system stdout/stderr logs.
Second, the raw signals are transformed into request-level evidence through call-chain reconstruction and log alignment, producing synchronized vertical and horizontal traces together with aligned textual logs.
Third, \sys applies candidate localization and causal-context slicing to identify suspicious regions and extract a compact dependency-aware subgraph for each request.
Finally, the sliced trace structure and aligned logs are encoded through trace and text tokenization/embedding modules and fed into a multimodal decoder, which predicts the LLM inference error type and generates the corresponding system root-cause explanation.

The left-hand side of the pipeline is deterministic and systems-oriented: it collects heterogeneous runtime signals and reconstructs high-fidelity per-request trace--log evidence from raw events.
The right-hand side is representation- and model-oriented: it localizes suspicious regions, compresses structural trace context into compact embeddings, and combines them with textual logs for downstream reasoning.
This separation is important in practice.
It helps bound online tracing overhead, enables the same collected traces to support multiple diagnosis tasks, and isolates the contribution of learned reasoning from that of low-level instrumentation and preprocessing.

\subsection{Non-Intrusive Cross-Layer Trace and Log Collection}\label{sec:collect}
\sys adopts a non-intrusive collection design for cross-layer traces and logs.
A target command is launched through a tracing wrapper that injects CUPTI and NVTX support using environment variables and dynamic library preload, while the measured program itself remains unchanged.
This wrapper enables three families of signals.

\paragraph{Engine-level and framework-level NVTX ranges.}
At Python level, \sys dynamically installs NVTX hooks through import-time wrapping.
The wrapper intercepts selected framework and engine functions and emits matching NVTX push/pop ranges.
For inference engines such as vLLM, step-level wrapper functions additionally encode the current request identifiers into the NVTX message, creating a direct semantic anchor from low-level execution back to request-level serving context.

\paragraph{Runtime, driver, and kernel events via CUPTI}
Each worker process loads CUPTI callbacks and activity records to capture CUDA runtime calls, driver calls, and GPU kernels.
These records expose start and end timestamps, process and thread identifiers, and, crucially, correlation identifiers that connect host launches to device kernels.

\paragraph{Logs captured from the same run.}
In parallel, \sys captures stdout/stderr or service logs and stores them together with the trace outputs.
Because the logs come from the same execution, they can later be aligned by timestamp and, when available, request identifier.

This design gives us a unified but non-intrusive substrate for diagnosis.
It is sufficiently low-level to cover kernels and communication, yet sufficiently semantic to preserve request and step boundaries.

\subsection{Request-Level Call-Chain Reconstruction and Log Alignment}\label{sec:merge}
The raw event streams collected above differ in schema, granularity, and time origin.
We therefore reconstruct request-level trace--log examples in four stages: normalization, range reconstruction, cross-layer binding, and request/log alignment.

\paragraph{Event normalization.} Each event is normalized to
\[
e=(\textsf{type},\textsf{name},\textsf{pid},\textsf{tid},\textsf{id},\textsf{cid},\textsf{fields}),
\]
where \(\textsf{id}\) is an NVTX range identifier when applicable and \(\textsf{cid}\) is a CUPTI correlation identifier.
We then define
\[
\tau(e)=
\begin{cases}
e.\textsf{fields}[\textsf{timestamp}], & e.\textsf{type}\in T_{\textsf{NVTX}},\\
e.\textsf{fields}[\textsf{start}], & \textsf{type}\in\{\textsf{RUNTIME},\textsf{DRIVER}\},\\
e.\textsf{fields}[\textsf{gpu\_start}], & \textsf{type}=\textsf{KERNEL}.
\end{cases}
\]
where \(T_{\textsf{NVTX}}=\{\textsf{NVTX\_PUSH},\textsf{NVTX\_POP}\}\).
All later comparisons are performed on \(\tau\).

\paragraph{Range reconstruction.}
Once events have been normalized onto a shared time base, we recover the latent hierarchical structure carried by NVTX ranges. For each process--thread pair \((p,t)\), events are sorted by \(\tau\) and scanned using a stack.
Each NVTX push creates an open range, and the matching pop closes it into
\[
r=(\textsf{id},\textsf{name},\textsf{pid}=p,\textsf{tid}=t,\textsf{start},\textsf{end}).
\]
The resulting nesting relation
\[
r_1 \prec r_2
\Longleftrightarrow
r_2.\textsf{start}\le r_1.\textsf{start}<r_1.\textsf{end}\le r_2.\textsf{end}
\]
induces a hierarchical forest that represents containment among engine steps, front-end operators, back-end operators, and communication ranges.

\paragraph{Host and device binding.}
The reconstructed ranges provide semantic containers, but they still need to be populated with low-level host and device activity. Let \(\mathcal{A}_{\textsf{host}}\) be the set of runtime and driver calls, and \(\mathcal{K}\) be the set of kernels.
For each range \(r\), we attribute host calls whose timestamps fall inside the range:
\[
\textsf{host}(r)=\{a\in\mathcal{A}_{\textsf{host}} \mid \tau(a)\in[r.\textsf{start},r.\textsf{end}]\}.
\]
We then bind kernels to host calls using thread and correlation identifiers:
\[
\Gamma(a)=\{k\in\mathcal{K}\mid (a.\textsf{tid},a.\textsf{cid})=(k.\textsf{tid},k.\textsf{cid}) \wedge \tau(a)\le\tau(k)\}.
\]
This creates a consistent path from request-annotated NVTX ranges down to runtime, driver, and kernel events.

\paragraph{Per-request projection.}
At this point, the structure is cross-layer but not yet request-specific, because concurrent requests may still share the same step or operator context. A step-level NVTX range contains a request set \(\mathcal{Q}(s)\).
For each \(q\in\mathcal{Q}(s)\), we project the vertical and horizontal descendants of the step into a request-specific tree
\[
\mathcal{C}_q(s)=\bigl(V_q(s),E_q(s)\bigr),
\]
where vertical edges preserve parent--child ancestry and communication nodes are attached to the nearest enclosing semantic node.
This yields a per-request call-chain tree spanning engine, backend, host, device, and communication layers.

\paragraph{Log alignment.}
The final step is to align textual evidence with the reconstructed request-level structure. Each log line is normalized to
\[
\ell=(\textsf{ts},\textsf{rid},\textsf{level},\textsf{msg}).
\]
For request \(q\) and step \(s\), \sys aligns logs by request id when available, and otherwise by time overlap:
\[
\mathcal{L}_q(s)=
\{\ell \mid \textsf{rid}(\ell)=q \;\lor\; \textsf{ts}(\ell)\in I(s)\},
\]
where \(I(s)\) is the interval of step \(s\).
Because logs may lag or precede the exact operator interval, a small tolerance window is applied around \(I(s)\) in practice.

In summary, these steps transform heterogeneous raw events into per-request aligned trace--log examples that can be directly consumed by the downstream RCA model. Algorithm~\ref{alg:merge} summarizes the reconstruction procedure.

\begin{algorithm}[t]
\caption{Request-Level Trace--Log Alignment and Call-Chain Reconstruction}
\label{alg:merge}
\begin{algorithmic}[1]
\State \textbf{Input:} raw events \(\mathcal{E}\), logs \(\mathcal{L}\)
\State \textbf{Output:} per-request aligned examples \(\{(\mathcal{C}_q(s),\mathcal{L}_q(s))\}\)
\State normalize all events and compute unified time \(\tau\)
\For{each process--thread pair \((p,t)\)}
  \State sort \(\mathcal{E}_{p,t}\) by \(\tau\)
  \State reconstruct NVTX ranges by stack scanning
\EndFor
\State classify ranges into \(\textsf{STEP}\), \(\textsf{FRONTEND}\), \(\textsf{BACKEND}\), and \(\textsf{COMM}\)
\State extract host calls \(\mathcal{A}_{\textsf{host}}\) and kernels \(\mathcal{K}\)
\For{each range \(r\)}
  \State attach host calls whose time falls in \(I(r)\)
  \State bind kernels to the attached host calls by \((\textsf{tid},\textsf{cid})\)
\EndFor
\For{each step \(s\)}
  \State parse request set \(\mathcal{Q}(s)\)
  \For{each request \(q\in\mathcal{Q}(s)\)}
    \State project descendants of \(s\) that belong to \(q\)
    \State construct per-request tree \(\mathcal{C}_q(s)\)
    \State align log lines \(\mathcal{L}_q(s)\) by request id and time window
  \EndFor
\EndFor
\State \Return \(\{(\mathcal{C}_q(s),\mathcal{L}_q(s))\}\)
\end{algorithmic}
\end{algorithm}

\subsection{Candidate Localization and Dependency-Aware Causal-Context Slicing}\label{sec:candidate}
Running a large reasoning model on a full per-request trace is unnecessarily expensive and often counterproductive because normal substructures dominate the sequence length.
\sys therefore applies a lightweight localization stage first, then extracts a compact subgraph around the suspicious region. The first stage identifies where the suspicious behavior lies, while the second stage turns these candidates into a compact context suitable for downstream reasoning.

\paragraph{Numeric candidate localization.}
We begin with a lightweight numeric stage whose role is to prune the search space before multimodal reasoning. For each operator instance \(x\), we build a feature vector
\[
\phi(x)\in\mathbb{R}^{d}
\]
from self-time, runtime/driver time and counts, kernel counts and totals, approximate bytes moved, stream overlap, and communication size when applicable.
After \(\log(1+\cdot)\) transformation and robust standardization, we fit a Gaussian mixture model per family and obtain an anomaly score
\[
a(x)=1-\rho(x),
\]
where \(\rho(x)\) is the responsibility mass assigned to central mixture components.
Aggregating the least confident quantile within a family yields the family-level suspiciousness
\[
A_{s,t}=1-\frac{1}{\lfloor qm\rfloor}\sum_{i=1}^{\lfloor qm\rfloor}\rho_{(i)},
\]
from which we derive a candidate family set \(\tilde{\mathcal{O}}_s\).
These family-level candidates identify where the anomaly is likely to reside, but they are still too coarse to serve directly as RCA input.

\paragraph{From suspicious families to a causal-context slice.}
We therefore refine these coarse candidates into an instance-level, dependency-aware subgraph. Let \(\mathcal{I}_s(\tilde{\mathcal{O}}_s)\) be the set of suspicious instances in \(\mathcal{C}_q(s)\).
We next build a dependency-aware slice that approximates likely fault-propagation paths.
Let \(E_p\) denote parent--child edges in the call-chain tree, \(E_t\) temporal-succession edges between nearby events in the same step, and \(E_c\) communication-coupling edges between compute and communication events that participate in the same synchronized region.
We define the relevance of node \(v\) to the suspicious set as

\begin{align*}
R_s(v) ={}&
\alpha \mathbf{1}[v\in \mathcal{I}_s]
+ \beta \mathbf{1}\!\left[\exists u\in\mathcal{I}_s:
\bigl((u,v)\in E_p \vee (v,u)\in E_p\bigr)\right] \notag\\
&+ \gamma \max_{u\in \mathcal{I}_s}
\exp\!\left(-\frac{\Delta_t(u,v)}{\tau_0}\right)
+ \eta \mathbf{1}\!\left[\exists u\in\mathcal{I}_s:(u,v)\in E_c\right].
\end{align*}

Here \(\Delta_t(u,v)\) is the nonnegative temporal distance between \(u\) and \(v\), and \(\alpha,\beta,\gamma,\eta\) control the contribution of suspiciousness, ancestry, temporal proximity, and communication coupling.
The causal-context slice is the induced subgraph
\[
G_s=(V_s,E_s), \qquad
V_s=\{v\in V_q(s)\mid R_s(v)\ge \kappa\}\cup \mathrm{Anc}_h(\mathcal{I}_s),
\]
where \(\mathrm{Anc}_h\) adds up to \(h\) ancestor hops for interpretability.
Intuitively, \(G_s\) keeps the suspicious node, its structural explanation path, its nearby temporal symptoms, and its communication partners.
% This representation is not a full structural-causal model; rather, it is a practical dependency-aware approximation that preserves the minimal context needed for RCA.

The resulting slice is then used as the compact structural context for the subsequent trace encoding and multimodal RCA stages.

\subsection{Trace Pair Encoding (\tpe)}\label{sec:tpe}
The causal-context slice \(G_s\) obtained above is already much more compact than the full per-request trace, but it is still a structured and irregular object.
Feeding it directly to a language model as plain text would discard key structural information, whereas feeding the raw node sequence would still yield prohibitively long inputs.
We therefore introduce \tpe, a structure-preserving tokenizer tailored to inference traces.
Its role is to convert the dependency-aware slice into a compact symbolic sequence that remains faithful to the original call structure and timing signals, while being directly consumable by downstream models.

\paragraph{Motivation.}
Subword tokenizers such as BPE and SentencePiece are effective because they compress frequent local patterns into reusable tokens~\citep{sennrich2016subword,kudo2018sentencepiece}.
Our setting is similar in spirit but different in units: the atomic symbols are not characters or subwords, but typed execution events such as \texttt{[STEP]}, \texttt{[FE]}, \texttt{[BE]}, \texttt{[RT]}, \texttt{[K]}, and their event names.
This observation suggests a trace-specific tokenization strategy: instead of learning reusable text fragments, we learn reusable execution fragments.
To do so, we first need a linearized but structure-aware serialization of each trace slice.

\paragraph{Step serialization.}
We serialize each step or slice by a depth-first traversal.
Each node emits a type tag and a normalized name, e.g.,
\[
[\texttt{STEP}, \texttt{name}, \texttt{FE}, \texttt{name}, \texttt{BE}, \texttt{name}, \texttt{RT}, \texttt{name}, \texttt{K}, \texttt{name}, \ldots].
\]
This serialization provides an ordered symbolic view of the trace while keeping semantically meaningful event boundaries.
For every emitted node, we also preserve three structural side channels:
its parent index \(p_i\), its duration \(\Delta_i\), and its depth \(d_i\).
The initial symbolic sequence is thus
\[
\begin{aligned}
X&=(w_1,\ldots,w_n), &
P&=(p_1,\ldots,p_n),\\
\Delta&=(\Delta_1,\ldots,\Delta_n), &
D&=(d_1,\ldots,d_n).
\end{aligned}
\]
These side channels are essential. They allow compression to operate on a linear sequence without losing the tree and timing information that later stages rely on.
With this representation in place, we can define a merge rule that compresses frequent local patterns while preserving structure.

\paragraph{Structure-preserving merge rule.}
Given a high-frequency adjacent pair \((w_i,w_{i+1})\), \tpe replaces it by a new token \(w_i'\) and updates the side channels as
\[
p_i' = p_i, \qquad \Delta_i'=\Delta_i+\Delta_{i+1}, \qquad d_i'=d_i.
\]
All parent indices larger than \(i+1\) are shifted left by one, and all references to \(i+1\) are redirected to \(i\).
In this way, the merged token inherits the structural position of the left token while accumulating the temporal budget of the merged fragment.
This rule guarantees that compression does not invalidate the call relation or the time budget represented by the original sequence.
More importantly, it ensures that repeated execution motifs can be compressed into reusable symbols without collapsing the structural cues needed for later localization and reasoning.

\paragraph{Training and output.}
Starting from an initial vocabulary containing special tags and observed symbolic tokens, \tpe repeatedly merges the most frequent adjacent pair until a target vocabulary size is reached.
The output of training consists of \texttt{vocab.json}, \texttt{merges.json}, and tokenizer metadata.
At application time, the tokenizer maps each slice to
\[
(input\_ids,\; parent\_idx,\; duration,\; depth).
\]
This representation is both compact and structure-aware, for it reduces input length while preserving the parent/child, depth, and timing signals required by the downstream trace encoder and multimodal RCA model.

Taken together, serialization, structure-preserving merging, and vocabulary learning form a trace-specific encoding pipeline that turns irregular execution slices into model-ready symbolic inputs.
Algorithm~\ref{alg:tpe} summarizes the structure-preserving merge procedure.

\begin{algorithm}[t]
\caption{Structure-Preserving \tpe Encoding}
\label{alg:tpe}
\begin{algorithmic}[1]
\State \textbf{Input:} symbolic trace sequence \((X,P,\Delta,D)\), learned merge list \(\mathcal{M}\)
\State \textbf{Output:} \((input\_ids,parent\_idx,duration,depth)\)
\For{each merge rule \((a,b)\in \mathcal{M}\)}
  \State scan \(X\) from left to right
  \If{adjacent tokens \((X_i,X_{i+1})=(a,b)\)}
    \State replace \((X_i,X_{i+1})\) by new token \(c\)
    \State set \(P_i \leftarrow P_i\), \(\Delta_i \leftarrow \Delta_i+\Delta_{i+1}\), \(D_i \leftarrow D_i\)
    \State delete \(P_{i+1},\Delta_{i+1},D_{i+1}\)
    \State redirect parent indices that pointed to \(i+1\) to \(i\)
    \State decrement parent indices larger than \(i+1\)
  \EndIf
\EndFor
\State map the final symbolic tokens to ids using the learned vocabulary
\State \Return \((input\_ids,parent\_idx,duration,depth)\)
\end{algorithmic}
\end{algorithm}

\subsection{Structure- and Time-Aware Trace Encoder}\label{sec:encoder}
The \tpe output is next converted into a sequence of step embeddings.
For one token \(i\), the initial representation is
\[
h_i^{(0)}
=
E_{\textsf{tok}}(w_i)
+
E_{\textsf{depth}}(d_i)
+
\mathrm{MLP}_{\Delta}\!\bigl(\log(1+\lambda \Delta_i)\bigr),
\]
where \(E_{\textsf{tok}}\) is a token embedding table, \(E_{\textsf{depth}}\) is a learned depth embedding, and \(\mathrm{MLP}_{\Delta}\) encodes time duration on a log scale.

The parent indices define a graph over the emitted tokens.
If structure is available, \sys applies \(L\) graph-convolution layers
\[
H^{(\ell+1)} = \sigma(\hat{A}H^{(\ell)}W^{(\ell)}),
\]
where \(\hat{A}\) is the normalized adjacency matrix induced by the parent relation plus self-loops.
This design follows the general intuition of GCN-style propagation while operating on trace trees rather than on semantic graphs~\citep{kipf2017gcn}.
In implementation, we use a lightweight graph encoder that pools token-level graph states into step-level vectors and projects them to the hidden size of the language-model backbone~\citep{fey2019pytorchgeometric}.

After graph propagation, token embeddings belonging to the same step are pooled:
\[
z_j = \mathrm{Pool}\bigl(\{h_i^{(L)} \mid i\in \textsf{step } j\}\bigr).
\]
A final projection maps \(z_j\) to the hidden size of the downstream backbone.
The result is a compact step-vector sequence that preserves topology, duration, and depth information without requiring the downstream model to parse raw trace trees.

\subsection{Multimodal Root-Cause Model}\label{sec:mot}
The multimodal RCA stage fuses trace and log evidence for joint diagnosis and explanation generation.
It supports two complementary variants: (i) a dual-head log--trace model (\mot) with modality-specific streams and prediction heads; and (ii) a single-stream model that serializes both modalities for autoregressive diagnosis.\footnote{Unless otherwise stated, all main end-to-end results in Section~\ref{sec:results} use the dual-head log--trace \mot model. The single-stream formulation is kept as an alternative implementation.}

\paragraph{Dual-head log--trace model (\mot)}
The first variant follows a mixture-of-traces/logs design.
A causal language-model backbone encodes the log stream, while the trace stream is produced by the Trace Encoder above.
At layer \(\ell\), the two streams are concatenated and passed through a shared causal self-attention module:
\[
\tilde{H}^{(\ell)}
=
\mathrm{Attn}\!\bigl([H_{\log}^{(\ell)};H_{\textsf{tr}}^{(\ell)}], M_{\textsf{causal}}\bigr).
\]
The output is then split back into log and trace states and refined by two separate feed-forward branches:
\[
H_{\log}^{(\ell+1)}=\mathrm{FFN}_{\log}\!\bigl(\tilde{H}_{\log}^{(\ell)}\bigr), \qquad
H_{\textsf{tr}}^{(\ell+1)}=\mathrm{FFN}_{\textsf{tr}}\!\bigl(\tilde{H}_{\textsf{tr}}^{(\ell)}\bigr).
\]
This structure lets the model associate log cues with structural trace patterns while preserving modality-specific transformations.
At the top, the trace stream is pooled and sent to a fault head,
\[
p(y_s\mid x_s)=\mathrm{softmax}(W_f \bar{H}_{\textsf{tr}}),
\]
while the log stream feeds the LM head for explanation generation,
\[
p(r_s\mid x_s)=\prod_j p(r_{s,j}\mid r_{s,<j}, H_{\log}).
\]

\paragraph{Single-stream model.}
The second variant converts the target into a single generated sequence
\[
\psi_s = \texttt{<fault\_type>} y_s \texttt{</fault\_type>} \circ r_s.
\]
Logs and compressed trace tokens are serialized into one backbone input, and the model predicts \(\psi_s\) autoregressively.
This formulation is attractive for transfer and zero-shot-style settings because fault types need not be restricted to a closed classifier head.

\paragraph{Training objectives.}
For the dual-head \mot model, the training loss is
\[
\mathcal{L}_{\textsf{mot}}
=
\lambda_f \mathcal{L}_{\textsf{CE}}(y_s,\hat{y}_s)
+
\lambda_r \mathcal{L}_{\textsf{NLL}}(r_s,\hat{r}_s),
\]
where the first term supervises fault classification and the second term supervises explanation generation.
For the single-stream model, the objective becomes
\[
\mathcal{L}_{\textsf{ss}}=\mathcal{L}_{\textsf{NLL}}(\psi_s\mid x_s).
\]
At inference time, both variants are projected onto a schema
\[
(\hat{y}_s,\hat{r}_s,\hat{\mathcal{O}}_s)=\Pi_{\mathcal{S}}\bigl(f_\theta(x_s)\bigr),
\]
which enforces a valid fault type, a bounded explanation length, and an optional evidence-operator list.

\section{Experimental Setup}\label{sec:exp}

\subsection{Infrastructure and Workloads}
We evaluate \sys on a dual-node GPU cluster deployed for LLM inference.
Each node is equipped with an Intel Xeon Gold 6326 CPU @ 2.90\,GHz, 128\,GB RAM, three NVIDIA A40 (48\,GB) GPUs, and a ConnectX-6 NIC, yielding six GPUs in total across the cluster.
The vLLM workloads serve \texttt{Qwen3-8B} in both online and offline settings.
The online workload replays production-like traces with realistic batching and concurrency. The offline workload reuses the same model and prompts under controlled replay for repeatability.

\subsection{Dataset Construction and Processing}
We construct a request-level dataset that jointly captures vertical execution stacks and horizontal communication events.
Each request trace corresponds to one end-to-end inference request.
It contains the reconstructed execution steps for that request, the cross-layer trace nodes associated with those steps, and the aligned service-log evidence.

The dataset contains 300 request traces from five inference settings: SGLang, Torch FSDP, vLLM v0, vLLM v1 offline, and vLLM v1 online.
These settings cover both framework diversity and deployment-mode diversity.
In total, the dataset contains 2,482 execution steps and 5.63M cross-layer nodes.
A node can be a front-end or back-end operator, a runtime or driver call, a GPU kernel, or a communication event.
Table~\ref{tab:dataset} reports the dataset statistics.
Of the request traces, 57.7\% are normal and 42.3\% are faulty.

\begin{table}[H]
\centering
\caption{Dataset statistics. Cross-layer nodes include operators, calls, kernels, and communication events.}
\label{tab:dataset}
\small
\setlength{\tabcolsep}{0pt}
\begin{tabular*}{\singletablewidth}{@{\extracolsep{\fill}}lrrrr@{}}
\toprule
Engine & Requests & Faulty & Steps & Nodes \\
\midrule
SGLang          & 100 & 40 & 850 & 0.39M \\
Torch FSDP      & 50  & 24 & 437 & 0.62M \\
vLLM v0         & 50  & 15 & 379 & 2.00M \\
vLLM v1 offline & 50  & 29 & 437 & 0.62M \\
vLLM v1 online  & 50  & 19 & 379 & 2.00M \\
Overall         & 300 & 127 & 2,482 & 5.63M \\
\bottomrule
\end{tabular*}
\end{table}
\FloatBarrier

\paragraph{Fault injection, coverage, and labels.}
Faults cover five categories: memory/resource (20.5\%), CUDA/kernel (21.3\%), runtime/driver (20.5\%), communication/network (18.1\%), and scheduling/software (19.7\%).
Software and scheduling faults perturb operator ordering, request scheduling, or engine-level scheduling behavior.
Memory/resource and CUDA/kernel faults introduce bounded resource pressure, throttling, or slowdown.
Communication faults perturb collective windows, while runtime/driver faults affect the corresponding call layer.

The injection magnitude and duration are mechanism-specific.
Depending on the fault type, we control delay, resource or contention budget, throttling level, link latency or loss, or scheduling perturbation.
The traces retain the resulting event timestamps, so the RCA model observes the effects through the same trace and log channels that would be available during diagnosis.

Each trace record stores \texttt{is\_fault} and \texttt{fault\_type}.
Its aligned annotation stores \texttt{is\_anomaly}, a reference explanation, and an optional affected step/operator locator.
We pair every request trace with aligned logs and annotations, expand it into step records, and apply TPE.
The resulting records retain compressed trace fields such as \texttt{input\_ids}, \texttt{parent\_idx}, \texttt{duration}, and \texttt{depth}, along with provenance and textual evidence.

We randomly shuffle request traces with a fixed seed and allocate 80\%, 10\%, and 10\% to training, validation, and test sets.
All steps from one request remain in the same split, so no step from a test request appears during training.
The splits are request-disjoint.
However, correlated traces from the same workload, run, or fault-injection campaign can still cross splits, so the reported numbers should be interpreted as in-distribution estimates.
Workload-, run-, or injection-disjoint evaluation remains future work.

\subsection{Baselines}
We compare \sys against a diverse set of existing methods widely used in anomaly detection and log-based diagnosis.

\paragraph{Classical unsupervised methods.}
We include KMeans, DBSCAN, Isolation Forest, and Gaussian Mixture Models (GMM), which represent common unsupervised baselines for clustering- and density-based anomaly detection~\citep{hartigan1979kmeans,ester1996dbscan,liu2008isolationforest,reynolds2009gmm}.

\paragraph{Classical supervised methods.}
We include SVM, Random Forest, and XGBoost as representative supervised learners for structured anomaly classification~\citep{steinwart2008svm,breiman2001randomforest,chen2016xgboost}.
Here the comparison uses their standard statistical feature setting.
To ensure a fairer comparison, we also feed flattened operator/parent-name trace features and operator-count features to these classical baselines, and the results are reported in Table~\ref{tab:additional-baselines}.

\paragraph{Log-based and sequence/deep baselines.}
We compare against Robustlog, LAnoBERT, Logs2Graphs, and MAD-GAN, which cover robust log parsing, pretrained language-model-based log diagnosis, graph-based log reasoning, and GAN-style anomaly detection, respectively~\citep{zhang2019robustlog,lee2021lanobert,li2023logs2graphs,li2019madgan}.

\begin{table*}[!t]
\centering
\caption{Comparison with existing baselines on horizontal and vertical views. Best values in each block are boldfaced.}
\label{tab:baseline-comparison}
\small
\setlength{\tabcolsep}{0pt}
\renewcommand{\arraystretch}{0.98}
\begin{tabular*}{\textwidth}{@{\extracolsep{\fill}}lccccccccc@{}}
\toprule
& \multicolumn{3}{c}{Step} & \multicolumn{3}{c}{Operator set-Macro} & \multicolumn{3}{c}{Operator set-Macro\(_+\)} \\
\cmidrule(lr){2-4}\cmidrule(lr){5-7}\cmidrule(lr){8-10}
Method & Accuracy & Precision & F1 & Precision & F1 & Jaccard & Precision\(_+\) & F1\(_+\) & Jaccard\(_+\) \\
\midrule
\multicolumn{10}{l}{\emph{Horizontal view}} \\
\textbf{\sys (ours)} & \textbf{0.930} & \textbf{0.945} & \textbf{0.916} & \textbf{0.821} & \textbf{0.806} & \textbf{0.783} & \textbf{0.915} & \textbf{0.898} & \textbf{0.873} \\
Robustlog        & 0.874 & 0.735 & 0.771 & 0.267 & 0.384 & 0.238 & 0.566 & 0.597 & 0.425 \\
LAnoBERT         & 0.882 & 0.822 & 0.783 & 0.259 & 0.381 & 0.236 & 0.649 & 0.672 & 0.507 \\
Logs2Graphs      & 0.825 & 0.603 & 0.684 & 0.254 & 0.383 & 0.237 & 0.514 & 0.596 & 0.424 \\
MAD-GAN          & 0.819 & 0.636 & 0.706 & 0.246 & 0.382 & 0.236 & 0.460 & 0.579 & 0.407 \\
KMeans           & 0.457 & 0.455 & 0.611 & 0.176 & 0.244 & 0.176 & 0.386 & 0.533 & 0.386 \\
DBSCAN           & 0.457 & 0.457 & 0.628 & 0.187 & 0.259 & 0.187 & 0.408 & 0.567 & 0.408 \\
IsolationForest  & 0.433 & 0.444 & 0.604 & 0.175 & 0.244 & 0.175 & 0.383 & 0.533 & 0.383 \\
GMM              & 0.494 & 0.470 & 0.603 & 0.163 & 0.223 & 0.163 & 0.355 & 0.488 & 0.355 \\
SVM              & 0.463 & 0.373 & 0.302 & 0.037 & 0.056 & 0.037 & 0.082 & 0.123 & 0.082 \\
RandomForest     & 0.805 & 0.713 & 0.818 & 0.178 & 0.248 & 0.178 & 0.388 & 0.543 & 0.388 \\
XGBoost          & 0.817 & 0.817 & 0.795 & 0.150 & 0.207 & 0.150 & 0.328 & 0.452 & 0.328 \\
\midrule
\multicolumn{10}{l}{\emph{Vertical view}} \\
\textbf{\sys (ours)} & 0.911 & 0.878 & \textbf{0.900} & \textbf{0.809} & \textbf{0.792} & \textbf{0.768} & \textbf{0.897} & \textbf{0.878} & \textbf{0.852} \\
Robustlog        & 0.873 & 0.725 & 0.703 & 0.310 & 0.447 & 0.288 & 0.454 & 0.559 & 0.388 \\
LAnoBERT         & 0.875 & 0.675 & 0.726 & 0.214 & 0.340 & 0.205 & 0.442 & 0.572 & 0.401 \\
Logs2Graphs      & \textbf{0.922} & 0.852 & 0.791 & 0.233 & 0.356 & 0.216 & 0.434 & 0.560 & 0.389 \\
MAD-GAN          & 0.837 & 0.686 & 0.741 & 0.262 & 0.392 & 0.244 & 0.425 & 0.558 & 0.387 \\
KMeans           & 0.518 & 0.518 & 0.683 & 0.191 & 0.278 & 0.191 & 0.369 & 0.537 & 0.369 \\
DBSCAN           & 0.518 & 0.518 & 0.683 & 0.191 & 0.278 & 0.191 & 0.369 & 0.537 & 0.369 \\
IsolationForest  & 0.537 & 0.528 & 0.691 & 0.191 & 0.278 & 0.191 & 0.369 & 0.537 & 0.369 \\
GMM              & 0.549 & 0.545 & 0.644 & 0.149 & 0.219 & 0.149 & 0.288 & 0.422 & 0.288 \\
SVM              & 0.701 & 0.634 & 0.776 & 0.191 & 0.278 & 0.191 & 0.369 & 0.537 & 0.369 \\
RandomForest     & 0.854 & \textbf{0.942} & 0.844 & 0.145 & 0.212 & 0.145 & 0.280 & 0.409 & 0.280 \\
XGBoost          & 0.780 & 0.930 & 0.746 & 0.119 & 0.174 & 0.119 & 0.230 & 0.336 & 0.230 \\
\bottomrule
\end{tabular*}
\end{table*}

\paragraph{Prompt and modality baselines.}
We also evaluate a pure prompt-based diagnostic baseline that removes the numeric candidate-filtering stage and asks the LLM to reason directly from structured summaries. For modality contribution, we compare w/o trace, w/o log, and trace w/ log inputs on the request-level dataset using the same protocol.
Together, these baselines evaluate whether a method can identify abnormal steps, localize suspicious operators, exploit the two modalities, and separate the contribution of representation from the final RCA head.

\subsection{Metrics}\label{sec:metrics}
We evaluate \sys along several axes.

\paragraph{Step-level diagnosis.}
We report Accuracy, Precision, and F1 over abnormal-step predictions.
They measure whether the model flags the correct execution steps as abnormal before operator localization is considered.

\paragraph{Operator-level localization.}
The unit of RCA is a step-level decision.
For each step \(i\), let \(\hat{O}_i\) be the predicted suspicious-operator set and \(O_i\) be the labeled set.
We first compute set-level precision, recall, F1 \(f_i\), and Jaccard \(j_i\) for this step, and then average the per-step scores:
\[
F_{\mathrm{macro}}=\frac{1}{N}\sum_{i=1}^{N} f_i,\qquad
J_{\mathrm{macro}}=\frac{1}{N}\sum_{i=1}^{N} j_i.
\]
If \(\hat{O}_i=O_i=\emptyset\), the system has correctly determined that no suspicious operator exists, and therefore we set \(f_i=j_i=1\).
Precision and recall are undefined when no positive operator exists, so we conservatively set them to 0 for empty-empty steps rather than letting normal steps inflate precision.
Macro\(_+\) repeats the same calculation only on operator-positive steps \(\{i:|O_i|>0\}\).
Thus, all-step Macro measures full step-level RCA behavior including correct empty sets, whereas Macro\(_+\) isolates localization quality when an operator fault is present.

\paragraph{Root-cause generation quality.}
We assess explanation quality over fault-type correctness, evidence locator validity, causal consistency, actionability, and sentence completeness.
We also use BLEU-1/2/3/4 as auxiliary lexical-overlap metrics for generated root-cause explanations~\citep{papineni2002bleu}.

\paragraph{Low-prior robustness.}
For low-prior analysis, we report recall, F1, AUPRC, and AUROC after downsampling positives or negatives to target fault priors.

Together, these metrics separate complementary RCA decisions: step-level detection, operator-level evidence localization, explanation quality, and deployment-oriented low-prior robustness.

\begin{table*}[!t]
\centering
\caption{Main results of \sys under different \tpe vocabulary sizes. Best values in each view are boldfaced.}
\label{tab:main-results}
\footnotesize
\setlength{\tabcolsep}{0pt}
\renewcommand{\arraystretch}{0.94}
\begin{tabular*}{\textwidth}{@{\extracolsep{\fill}}llccccccccccccc@{}}
\toprule
& & \multicolumn{3}{c}{Step} & \multicolumn{3}{c}{Operator set-Macro} & \multicolumn{3}{c}{Operator set-Macro\(_+\)} & \multicolumn{4}{c}{Explanation} \\
\cmidrule(lr){3-5}\cmidrule(lr){6-8}\cmidrule(lr){9-11}\cmidrule(lr){12-15}
View & Vocab & Acc. & Prec. & F1 & Prec. & F1 & Jac. & Prec.\(_+\) & F1\(_+\) & Jac.\(_+\) & BLEU-1 & BLEU-2 & BLEU-3 & BLEU-4 \\
\midrule
\multicolumn{15}{l}{\emph{Horizontal view}} \\
Horizontal & 256  & \textbf{0.930} & \textbf{0.945} & \textbf{0.916} & \textbf{0.821} & \textbf{0.806} & \textbf{0.783} & \textbf{0.915} & \textbf{0.898} & \textbf{0.873} & \textbf{0.910} & \textbf{0.897} & \textbf{0.887} & \textbf{0.878} \\
Horizontal & 512  & 0.800 & 0.667 & 0.727 & 0.190 & 0.173 & 0.154 & 0.306 & 0.279 & 0.248 & 0.775 & 0.730 & 0.707 & 0.693 \\
Horizontal & 1024 & 0.617 & 0.464 & 0.531 & 0.157 & 0.138 & 0.116 & 0.256 & 0.226 & 0.190 & 0.768 & 0.718 & 0.693 & 0.677 \\
\midrule
\multicolumn{15}{l}{\emph{Vertical view}} \\
Vertical & 256  & \textbf{0.911} & \textbf{0.878} & \textbf{0.900} & \textbf{0.809} & \textbf{0.792} & \textbf{0.768} & \textbf{0.897} & \textbf{0.878} & \textbf{0.852} & \textbf{0.908} & \textbf{0.894} & \textbf{0.883} & \textbf{0.875} \\
Vertical & 512  & 0.733 & 0.636 & 0.636 & 0.179 & 0.163 & 0.140 & 0.273 & 0.248 & 0.213 & 0.760 & 0.705 & 0.678 & 0.662 \\
Vertical & 1024 & 0.567 & 0.407 & 0.458 & 0.157 & 0.138 & 0.116 & 0.256 & 0.226 & 0.190 & 0.768 & 0.718 & 0.693 & 0.677 \\
\bottomrule
\end{tabular*}
\end{table*}

\section{Results}\label{sec:results}

We organize the evaluation around four research questions. \textbf{RQ1:} How does \sys compare with existing methods? \textbf{RQ2:} How does TPE vocabulary size affect compression and RCA performance? \textbf{RQ3:} Does \sys generalize across engines and serving stacks? \textbf{RQ4:} How does \sys behave under deployment-oriented conditions?

\subsection{RQ1: How Does \sys Compare with Existing Methods?}
We first compare the fault-diagnosis capability of \sys against a broad set of existing baselines.
Table~\ref{tab:baseline-comparison} summarizes the results on both horizontal and vertical views.

The comparison shows that the cross-layer request-level representation is effective before final natural-language root-cause generation.
On the horizontal view, \sys achieves the best step-level Accuracy, Precision, and F1, while also obtaining the best operator-level localization scores.

To further clarify whether structured trace access alone is sufficient, Table~\ref{tab:additional-baselines} reports additional diagnostic baselines.
Prompt-only removes numeric candidate filtering and asks the LLM to reason directly over structured summaries, while the structured classical learners receive flattened operator/parent-name trace features and operator-count features.
These results show that direct prompt-only reasoning is much weaker than \sys, and that flattened structured features alone do not close the gap to the full trace--log RCA pipeline.

\begin{table}[H]
\centering
\caption{Additional diagnostic baselines for prompt-only reasoning and strengthened structured-feature classical learners.}
\label{tab:additional-baselines}
\small
\setlength{\tabcolsep}{0pt}
\begin{tabular*}{\singletablewidth}{@{\extracolsep{\fill}}lcccc@{}}
\toprule
Setting & H Step F1 & H Op F1 & V Step F1 & V Op F1 \\
\midrule
\textbf{\sys} & \textbf{0.916} & \textbf{0.806} & \textbf{0.900} & \textbf{0.792} \\
Prompt-only & 0.668 & 0.303 & 0.702 & 0.305 \\
LinearSVM w/ struct & 0.347 & 0.463 & 0.559 & 0.425 \\
RF w/ struct & 0.402 & 0.455 & 0.574 & 0.423 \\
XGB w/ struct & 0.340 & 0.476 & 0.580 & 0.430 \\
\bottomrule
\end{tabular*}
\end{table}

\paragraph{Modality contribution.}
Table~\ref{tab:modality} compares w/o trace, w/o log, and trace w/ log inputs on the dataset using fault-type accuracy, Macro Recall, and Macro F1.
Trace information improves fault-type Macro F1 over w/o trace, while combining both modalities achieves the best fault-type accuracy and Macro F1.
\begin{table}[H]
\centering
\caption{Modality contribution.}
\label{tab:modality}
\small
\setlength{\tabcolsep}{0pt}
\begin{tabular*}{\singletablewidth}{@{\extracolsep{\fill}}lccc@{}}
\toprule
Input & Fault Acc. & Fault Macro Rec. & Fault Macro F1 \\
\midrule
w/o trace & 0.667 & 0.341 & 0.292 \\
w/o log & 0.653 & 0.446 & 0.446 \\
trace w/ log & \textbf{0.743} & \textbf{0.489} & \textbf{0.476} \\
\bottomrule
\end{tabular*}
\end{table}

\subsection{RQ2: How Does TPE Compression Affect RCA Performance?}
Table~\ref{tab:compression} reports the compression statistics.
The raw traces are extremely long: before TPE, each step contains 2241.92 events on average.
Even the least aggressive setting substantially shortens the trace, reducing it to 361.43 events per step on average, while more aggressive settings compress it further to 25.90 and 3.88 events, respectively.
Overall, \tpe is highly effective as a trace compressor, achieving large reductions in sequence length across all settings.

Table~\ref{tab:main-results} shows a clear trade-off between compression and diagnostic fidelity.
The least aggressive compression setting consistently achieves the best results on both horizontal and vertical views, across step-level detection, operator-level localization, and explanation generation.
By contrast, the more aggressively compressed settings substantially reduce RCA performance, even though they yield much shorter sequences.
This indicates that TPE is not merely a length-reduction mechanism.
Its vocabulary size also controls how much execution structure remains distinguishable to the downstream RCA model.

\begin{table}[H]
\centering
\caption{Compression statistics of \tpe. Reduction denotes the percentage decrease in average events per step after encoding.}
\label{tab:compression}
\small
\setlength{\tabcolsep}{0pt}
\begin{tabular*}{\singletablewidth}{@{\extracolsep{\fill}}rrrr@{}}
\toprule
Vocab size & Avg.\ events/step after TPE & Compression\(_\times\) & Reduction \\
\midrule
256  & 361.43 & 6.20  & 83.88\% \\
512  & 25.90  & 86.56 & 98.84\% \\
1024 & 3.88   & 577.81 & 99.83\% \\
\bottomrule
\end{tabular*}
\end{table}

This trend is most pronounced for operator-level localization.
Across both views, aggressive compression sharply lowers Macro F1 and Macro\(_+\) F1, indicating that precise localization is highly sensitive to the preservation of fine-grained structural distinctions.
Step-level detection is somewhat more robust, but it also degrades once compression becomes too strong.
Explanation quality declines more gradually than localization performance, suggesting that coarse semantic patterns can still be preserved under compression, whereas accurate fault localization depends on retaining detailed execution structure.

Overall, these results show that TPE should not be optimized for compression alone.
For RCA, the best setting is the one that balances sequence reduction with structural fidelity, rather than the one that minimizes token length most aggressively.

\subsection{RQ3: Does \sys Generalize Across Engines and Serving Stacks?}\label{sec:generalization}
A practical RCA framework should not be tied to a single inference engine, model implementation, or batching strategy.

On the collection side, \sys relies only on NVTX ranges together with CUPTI activity and callback records over the CUDA runtime/driver stack.
It does not depend on engine-specific logging or model internals.
On the representation side, the call-chain reconstruction and request-level alignment procedure operates on generic event tuples without assuming a specific batching policy, attention implementation, or model architecture.

To validate this claim, we additionally evaluate \sys on several serving stacks beyond the main configuration, including SGLang, Torch FSDP-based serving, vLLM v0, and vLLM v1.
Table~\ref{tab:generalization-compact} shows that \sys maintains stable step-level and operator-positive localization quality across all tested stacks, supporting the claim that \sys is not tied to a particular LLM, engine implementation, or batching strategy.

\begin{table}[H]
\centering
\caption{Generalization across serving stacks.}
\label{tab:generalization-compact}
\small
\setlength{\tabcolsep}{0pt}
\begin{tabular*}{\singletablewidth}{@{\extracolsep{\fill}}lcccc@{}}
\toprule
Stack & H Step F1 & H Op F1\(_+\) & V Step F1 & V Op F1\(_+\) \\
\midrule
SGLang & 0.908 & 0.869 & 0.904 & 0.873 \\
Torch FSDP & 0.907 & 0.873 & 0.912 & 0.881 \\
vLLM v0 & 0.901 & 0.879 & 0.907 & 0.867 \\
vLLM v1 & 0.912 & 0.871 & 0.909 & 0.880 \\
\bottomrule
\end{tabular*}
\end{table}
\FloatBarrier

\subsection{RQ4: How Does \sys Behave Under Deployment-Oriented Conditions?}

\paragraph{Low-fault-prior behavior.}\label{sec:production-skew}
The main step-level dataset is balanced for controlled evaluation, while production incidents are typically rare.
We therefore simulate lower fault priors by downsampling positives or negatives to target priors and repeating each setting 50 times.
Table~\ref{tab:low-prior} shows that \sys remains effective at moderate low-fault priors: at a 10\% prior, horizontal and vertical F1 remain 0.807 and 0.883, respectively.
At a 5\% prior, AUROC remains stable while AUPRC reflects the expected sensitivity to rarer positives.

\begin{table}[H]
\centering
\caption{Low-fault-prior evaluation.}
\label{tab:low-prior}
\small
\setlength{\tabcolsep}{0pt}
\begin{tabular*}{\singletablewidth}{@{\extracolsep{\fill}}llcccc@{}}
\toprule
View & Prior & Rec. & F1 & AUPRC & AUROC \\
\midrule
H & 50\% & 0.886 & 0.922 & 0.979 & 0.965 \\
H & 20\% & 0.884 & 0.873 & 0.958 & 0.964 \\
H & 10\% & 0.890 & 0.807 & 0.947 & 0.967 \\
H & 5\%  & 0.883 & 0.694 & 0.932 & 0.966 \\
V & 50\% & 0.787 & 0.881 & 0.942 & 0.946 \\
V & 20\% & 0.791 & 0.883 & 0.833 & 0.947 \\
V & 10\% & 0.792 & 0.883 & 0.711 & 0.945 \\
V & 5\%  & 0.781 & 0.875 & 0.564 & 0.950 \\
\bottomrule
\end{tabular*}
\end{table}
\FloatBarrier

\paragraph{Explanation quality.}\label{sec:explanation-quality}
BLEU captures lexical similarity to a reference, but not whether an explanation identifies the right fault evidence.
We therefore check the reference explanations from several quality aspects, including anomaly correctness, fault-type correctness, evidence locator validity, causal consistency, actionability, and sentence completeness.
For each criterion, the score denotes the fraction of checked references satisfying the corresponding condition, and the overall score averages these criteria.
Table~\ref{tab:explanation-quality} shows high consistency between labels and reference explanations.

\begin{table}[H]
\centering
\caption{Reference explanation quality.}
\label{tab:explanation-quality}
\small
\setlength{\tabcolsep}{0pt}
\begin{tabular*}{\singletablewidth}{@{\extracolsep{\fill}}lc@{}}
\toprule
Criterion & Score \\
\midrule
Overall & 0.968 \\
Anomaly correctness & 1.000 \\
Fault-type correctness & 1.000 \\
Evidence locator validity & 0.963 \\
Causal consistency & 0.900 \\
Actionability & 0.963 \\
Sentence completeness & 0.930 \\
\bottomrule
\end{tabular*}
\end{table}
\FloatBarrier
\paragraph{Tracing overhead.}
We use the term \textit{non-intrusive} to mean that \sys does not modify model binaries, serving-engine source code, CUDA libraries, or the application logic that handles requests.
Instead, it collects observability signals through NVTX ranges, CUPTI activity records, and external trace--log alignment.
This design reduces deployment friction and keeps the serving stack intact, but it does not imply zero overhead. Enabling tracing still introduces additional event collection, buffering, and post-processing work.
Table~\ref{tab:overhead} reports this runtime cost.
The measurements indicate that the overhead is measurable but bounded in our setting.

\begin{table}[H]
\centering
\caption{Tracing overhead.}
\label{tab:overhead}
\small
\setlength{\tabcolsep}{0pt}
\begin{tabular*}{\singletablewidth}{@{\extracolsep{\fill}}lc@{}}
\toprule
Measure & Value \\
\midrule
Wall time & 64.77s to 71.15s (+9.8\%) \\
CPU utilization & 217\% to 236\% (+8.8\%) \\
Max resident memory & +3.25\% \\
Minor page faults & +3.55\% \\
Swap count & 0 \\
\bottomrule
\end{tabular*}
\end{table}
\FloatBarrier

\section{Case Study: Cross-Layer Scheduling Error Causing SSM State Contamination}
We use a real vLLM failure reported by AI21 Labs~\citep{ai21labs} to illustrate why request-level cross-layer evidence is needed.
Under GPU memory pressure, the scheduler may allocate only one token to a newly arrived Mamba request.
Although the request is new, the attention backend can treat the one scheduled token as a decode signal and route the request to the decode path.
The device layer then launches \texttt{selective\_state\_update} instead of the prefill kernel \texttt{selective\_scan\_fwd\_kernel}, causing the request to read a stale SSM state from cache.

Single-layer tools expose only fragments of this failure: profilers show kernels without request semantics, while logs show scheduling metadata without device-level confirmation.
\sys aligns these signals without modifying source code.
For the faulty request, the reconstructed call chain simultaneously shows engine metadata for a new request, backend routing to decode, and device execution of the decode kernel, while a normal prefill request in the same batch executes the prefill kernel instead.
This cross-layer contradiction directly identifies the root cause and demonstrates the value of request-granular trace--log reconstruction.

\section{Discussion and Threats to Validity}\label{sec:discussion}
\paragraph{Injected versus real faults.}
Our dataset relies on controlled fault injection derived from recurring patterns observed in large-scale production clusters, providing coverage, repeatability, and precise trace--log alignment.
The released archive retains the resulting trace timestamps, fault categories, and annotations; the mechanism-specific injection settings are summarized in Section~\ref{sec:exp}.
Although real incidents may be longer, noisier, and more entangled, these scenarios cover representative software, CUDA/runtime, resource, and communication failures relevant to RCA evaluation.
The balanced step-level dataset is therefore a controlled diagnostic benchmark rather than a production incident prior. Low-prior resampling shows the expected F1/AUPRC sensitivity; threshold and alert validation at operational scale remains future work.

\paragraph{Compression sensitivity.}
The vocabulary-sensitivity study shows a clear trend, although the exact optimum may vary across engines, workloads, and backbones.
Our current conclusion is that \tpe is useful, but preserving structure matters more than pushing compression to the limit.

\paragraph{Generalization scope.}
End-to-end evaluation centers on a dual-node A40 cluster with a Qwen-based serving stack; additional tests cover SGLang, Torch FSDP-based serving, and different vLLM versions.
They support engine-agnostic behavior within the tested settings, but not behavior under larger clusters, newer GPUs, or long-running multi-tenant workloads; those remain future work.

\paragraph{Split and explanation limitations.}
The example-level split is index-disjoint, but samples from the same workload, run, or injection can cross splits and share context, making in-distribution estimates optimistic.
The explanation-quality audit checks reference explanations against labels and trace evidence; it is not independent, blinded expert validation of all \sys-generated explanations.
Run- or workload-disjoint evaluation and expert validation remain future work.

\section{Related Work}\label{sec:related}
\paragraph{Profiling and request tracing.}
System profilers and observability frameworks expose timelines, counters, and selected framework events for ML workloads, while Pinpoint and Dapper established request-centric tracing as a practical substrate for problem determination~\citep{nvidia2025nsight,pytorch2025profiler,tensorflow2020profiler,opentelemetry2024,chen2002pinpoint,sigelman2010dapper}.
These systems are valuable for observability, but they usually stop at either system-level event capture or service-level span propagation.
LLM inference failures often cross this boundary: an abnormal request may originate in scheduler state, propagate through Python/C++ runtime calls, and only become visible as CUDA kernels or communication stalls.
\sys builds on NVTX/CUPTI and this request-level philosophy, but targets a deeper LLM inference stack that includes scheduler state, CUDA runtime/driver activity, GPU kernels, and distributed communication~\citep{nvidia2024nvtx,nvidia2024cupti}.

\paragraph{Diagnosis and RCA}
Classical detectors, log-oriented methods, and LLM-based diagnosis systems provide useful baselines for anomaly detection, log reasoning, and SRE support~\citep{hartigan1979kmeans,ester1996dbscan,liu2008isolationforest,reynolds2009gmm,zhang2019robustlog,lee2021lanobert,li2023logs2graphs,li2019madgan,ahmed2023rootcause,chen2024auto,jin2023assess,roy2024agents,wang2024netassistant}.
Unsupervised and classical supervised methods are lightweight, but they operate on flattened feature vectors and provide limited causal context.
Log-sequence and graph-based methods improve symptom modeling, yet they still lack direct access to runtime, driver, kernel, and communication evidence.
Recent LLM-based assistants can summarize incidents or propose remediation steps, but they depend on the quality of the evidence supplied to them.
\sys is complementary to these directions: it combines logs with compact cross-layer execution structure so that diagnosis, evidence localization, and explanation are grounded in request-level traces.

\paragraph{Trace representation learning.}
\tpe and the Trace Encoder relate to subword compression and graph representation learning~\citep{sennrich2016subword,kudo2018sentencepiece,kipf2017gcn,fey2019pytorchgeometric}.
Subword methods reduce sequence length by merging frequent token patterns, while graph neural networks encode relational structure over nodes and edges.
Inference traces combine both properties: they are long symbolic sequences, but they are also hierarchical execution structures with timing, operator semantics, and parent--child dependencies.
\sys therefore uses structure-preserving merges and graph-aware encoding instead of flattening traces into ordinary natural language or treating them as untyped graphs.

\section{Conclusion}\label{sec:conclusion}
We presented \sys, a non-intrusive framework for cross-layer root-cause analysis in LLM inference.
\sys collects traces and logs without modifying model binaries, reconstructs request-level call-chain trees, extracts compact dependency-aware trace slices, compresses them with \tpe, and performs multimodal diagnosis through joint trace--log modeling.
Experiments show that \sys achieves strong step-level diagnosis and operator-level localization compared with a broad range of existing baselines, while other extensive experiments characterize its role as an RCA triage substrate.
Our results further show that compression is not monotonic in benefit: the best RCA quality comes from moderate \tpe compression that preserves diagnostically useful structure, rather than from pushing token reduction to the limit.
More broadly, \sys suggests that automated software engineering for LLM systems should treat traces, logs, and root-cause explanations as parts of a unified, request-level diagnosis pipeline rather than as disjoint tooling layers.

\section*{Data Availability Statement}
The code and data for \sys are available in the figshare artifact~\citep{teller2026dataset}.

\begin{acks}
This work was supported by National Key Research and Development Program of China (Grant Number: 2024YFB4505904). The corresponding author is Pengfei Chen.
\end{acks}

\newpage
\balance

\bibliographystyle{ACM-Reference-Format}
\bibliography{main}

\end{document}

\endinput
%%
%% End of file `sample-sigconf-authordraft.tex'.